\documentclass[aps,prd,onecolumn,preprint,preprintnumbers,nofootinbib,superscriptaddress,floatfix,longbibliography]{revtex4-1}
\pdfoutput=1
\usepackage{graphicx}
\usepackage{epstopdf}
\usepackage{mathrsfs}
\usepackage{amssymb}
\usepackage{verbatim}
\usepackage{color}
\usepackage{multirow}
\usepackage{stackrel}
\usepackage[pdfencoding=auto]{hyperref}

\usepackage{amsmath}

\DeclareMathOperator*{\argmin}{arg\,min}

\newcommand{\f}{\mathbf{f}}

\newcommand{\beq}{\begin{equation}}
\newcommand{\eeq}{\end{equation}}
\newcommand{\ga}{\lower.7ex\hbox{$\;\stackrel{\textstyle>}{\sim}\;$}}
\newcommand{\la}{\lower.7ex\hbox{$\;\stackrel{\textstyle<}{\sim}\;$}}

\usepackage{hyperref}
\usepackage{amsmath,bm}
\usepackage{physics}
\allowdisplaybreaks

\onecolumngrid

\hypersetup{
    colorlinks = true,
    citecolor = {blue},
    linkcolor = {blue},
    urlcolor = {blue},
}

\begin{document}

\title{Discovering Sparse Representations of Lie Groups\texorpdfstring{\\}{} with Machine Learning}

\author{Roy T.~Forestano}
\email{roy.forestano@ufl.edu}
\affiliation{Institute for Fundamental Theory, Physics Department, University of Florida, Gainesville, FL 32611, USA}

\author{Konstantin T.~Matchev}
\email{matchev@ufl.edu}
\affiliation{Institute for Fundamental Theory, Physics Department, University of Florida, Gainesville, FL 32611, USA}

\author{Katia Matcheva}
\email{matcheva@ufl.edu}
\affiliation{Institute for Fundamental Theory, Physics Department, University of Florida, Gainesville, FL 32611, USA}

\author{Alexander Roman}
\email{alexroman@ufl.edu}
\affiliation{Institute for Fundamental Theory, Physics Department, University of Florida, Gainesville, FL 32611, USA}

\author{Eyup B.~Unlu}
\email{eyup.unlu@ufl.edu}
\affiliation{Institute for Fundamental Theory, Physics Department, University of Florida, Gainesville, FL 32611, USA}

\author{Sarunas~Verner}
\email{verner.s@ufl.edu}
\affiliation{Institute for Fundamental Theory, Physics Department, University of Florida, Gainesville, FL 32611, USA}

\begin{abstract}
Recent work has used deep learning to derive symmetry transformations, which preserve conserved quantities, and to obtain the corresponding algebras of generators. In this letter, we extend this technique to derive sparse representations of arbitrary Lie algebras. We show that our method reproduces the canonical (sparse) representations of the generators of the Lorentz group, as well as the $U(n)$ and $SU(n)$ families of Lie groups. This approach is completely general and can be used to find the infinitesimal generators for any Lie group.
\end{abstract}

\date{February 10, 2023}

\maketitle

\section{Introduction}

The transformative advances in theoretical physics at the turn of the 20${}^{\rm th}$ century came with the realization that symmetry is a primary feature of nature that constrains the allowable dynamical laws \cite{Gross1996}. By Noether's theorem \cite{Noether1918}, the presence of a continuous symmetry in the problem implies a conservation law which is universally applicable and indispensable in understanding the system's behavior and evolution. Symmetries can be found at all scales --- from the microscopic description of subatomic particles in the Standard Model to the large-scale structure of the Universe.  In particle physics, symmetries provide an organizing principle behind the observed particle zoo and its interactions, and guide model-builders in the search for viable extensions of the Standard Model \cite{Csaki:2018muy}. 

The formal study of symmetries has historically been done within the framework of group theory \cite{wigner1959group}. While virtually all types of classical Lie groups find some applications in various physics subfields, the most commonly encountered Lie groups in particle physics are the special orthogonal groups $SO(n)$ and the special unitary groups $SU(n)$. In particular, the rotation group $SO(3)$ reflects the isotropy of space and all fundamental laws of motion, the Lorentz group $SO(1,3)$ is the symmetry group of spacetime in special relativity, and $SU(2)$ and $SU(3)$ are the Standard Model gauge groups responsible for the weak and strong interactions, respectively. An advanced course in group theory is a necessary tool in the arsenal of the aspiring physics graduate student. 

Recently there has been increased interest in the application of machine learning (ML) for the discovery and identification of symmetries in data \cite{Iten1807.10300,Wetzel:2020jan,Krippendorf:2020gny,Liu:2020omw,Barenboim:2021vzh,Dillon:2021gag,Liu:2021azq,Desai:2021wbb,Craven:2021ems,Moskalev2210.04345,Forestano:2023fpj,Roman:2023ypv}. Relevant applications of ML to group theory include computing tensor products and branching rules of irreducible representations of Lie groups \cite{Chen:2020jjw}, and testing and/or deriving Lie group generators of a symmetry present in the data \cite{Liu:2021azq,Moskalev2210.04345,Forestano:2023fpj,Roman:2023ypv}. In this letter we extend the method of \cite{Forestano:2023fpj,Roman:2023ypv} to derive {\em sparse} representations of arbitrary Lie algebras. We shall demonstrate that, with the appropriate loss function modifications discussed in Section~\ref{sec:loss}, one can use machine learning to derive from first principles the canonical (sparse) form of the generators for the Lorentz group (Section~\ref{sec:lorentz_group}), as well as for the $U(n)$ and $SU(n)$ families (Section~\ref{sec:SUN}). As emphasized in the summary Section~\ref{sec:summary}, the described method is completely general and can be applied to {\em any} other classical Lie group.

\section{Method}
\label{sec:loss}

We begin by briefly summarizing the symmetry finding procedure outlined in Ref.~\cite{Forestano:2023fpj}. The classical groups are the linear groups of transformations over the reals $\mathbb R$, the complex numbers $\mathbb C$, and quaternions $\mathbb  H$. Therefore, a symmetry transformation acts on a feature vector ${\mathbf x}\equiv \{x^{(1)}, x^{(2)},\ldots, x^{(n)}  \}$, where ${\mathbf x}\in \mathbb R^n$ and ${\mathbf x}\in \mathbb C^n$ in our $SO(n)$ and $SU(n)$ examples, respectively. In order to capture the effect of a group transformation on ${\mathbb R}^n$ or ${\mathbb C}^n$, we consider a representative set of $m$ points $\left\{\mathbf{x}\right\} \equiv \left\{\mathbf{x}_1,\mathbf{x}_2,\ldots,\mathbf{x}_m\right\}$ sampled from some finite domain. The choice of a sampling distribution and/or the size and location of the domain is inconsequential. For definiteness we use a standard normal distribution with $m=100$. 

The classical groups are defined in terms of polynomial invariants over their respective fields. For example, $O(n)$ preserves the values of the polynomial oracle
\beq
\varphi_{O}(\mathbf{x}) 
\; \equiv \; |\mathbf{x}|^2 = \sum_{j=1}^n [x^{(j)}]^2, 
\quad 
x^{(j)}\in \mathbb R,
\label{oracle:O}
\eeq
the Lorentz group in $n=4$ dimensions preserves 
\beq
\varphi_{L}(\mathbf{x}) 
\; \equiv \; 
\bigl(x^{(1)}\bigr)^2 -
\bigl(x^{(2)}\bigr)^2 -
\bigl(x^{(3)}\bigr)^2 -
\bigl(x^{(4)}\bigr)^2,
\quad 
x^{(j)}\in \mathbb R,
\label{oracle:L}
\eeq
while $U(n)$ preserves
\beq
\varphi_{U}(\mathbf{x}) 
\; \equiv \; \sum_{j=1}^n \bigl(x^{(j)}\bigl)^\ast x^{(j)}, 
\quad 
x^{(j)}\in \mathbb C.
\label{oracle:U}
\eeq

A symmetry transformation $\f$ is a map $\mathbf{x}' = \f(\mathbf{x})$ which preserves the respective oracle 
(\ref{oracle:O}-\ref{oracle:U}) everywhere, or in our case, for each of the sampled $m$ points:
\beq
\varphi(\mathbf{x}'_i) \; \equiv \; \varphi(\f(\mathbf{x}_i)) = \varphi(\mathbf{x}_i), \quad \forall i \,= \,1,2,\ldots,m \, .
\eeq
In order to focus on the {\em generators} of the group of symmetry transformations, $\f$ is linearized by considering infinitesimal transformations $\delta{\f}$ in the vicinity of the identity transformation $\mathbb{I}$:
\beq
\delta{\f} \; \equiv \; \mathbb{I} + \varepsilon \, {\mathbb G} \, ,
\label{eq:deltaf}
\eeq
where $\varepsilon$ is an infinitesimal parameter and ${\mathbb G}$ is a $n\times n$ matrix. Upon training with a suitable loss function $L({\mathbb G}, \{\mathbf x\})$, the components of ${\mathbb G}$ are driven to their {\em trained} values, thus producing one symmetry generator~\cite{Forestano:2023fpj}
\beq
\mathbb{J} \; \equiv \;  \argmin_{\mathbb G}
\Bigl(L({\mathbb G}, \{\mathbf x\}) \Bigr).
\eeq
By repeating the above procedure $N_g$ times under different initial conditions ${\mathbb G}_0$ for the candidate generator components, or for different random seeds and hyperparameters, one obtains a whole set of $N_g$ generators $\mathbb{J}_\alpha$, $\alpha=1,2,\ldots,N_g$. The loss function is chosen to ensure that the obtained generators have the following properties:
\begin{enumerate}
\item {\it Invariance}, i.e., preserving the oracle values for all sampled datapoints $\left\{\mathbf{x}\right\}$ under the set $\{\mathbb G\}$ of all candidate transformations $\mathbb G_\alpha$, $\alpha=1,2,\ldots,N_g$:
\beq
    L_\text{inv}\bigl(\{\mathbb G\}, \{\mathbf x\}\bigl) = \frac{1}{m\varepsilon^2}\sum_{\alpha=1}^{N_g}\sum_{i=1}^m \Bigl[ \varphi\bigl(\mathbf{x}_i + \varepsilon {\mathbb G_\alpha}\cdot\mathbf{x}_i\bigr)-\varphi(\mathbf{x}_i) \Bigr]^2 \, ,
\label{eq:LossInvariance}    
\eeq
where ``$\cdot$" denotes ordinary tensor multiplication.
\item {\it Orthogonality}, which guarantees that the obtained generators are distinct: 
\beq
 L_\text{ortho}\bigl(\{\mathbb G\} \bigr) 
 = \Bigl[
 \sum_{ \alpha=1}^{N_g-1} \sum_{\substack{\beta=\alpha+1 }}^{N_g} \text{Tr}\bigl({\mathbb G}_\alpha\cdot {\mathbb G}_\beta^T\bigr)\Bigr]^2.
\label{eq:LossOrthogonality}
\eeq
\item {\it Normalization}, ensuring that the transformation is not trivial:
\beq
L_\text{norm}\bigl(\{\mathbb G\}\bigr) = \sum_{ \alpha=1}^{N_g} \Bigl[ \text{Tr}\Bigl({\mathbb G}_\alpha\cdot {\mathbb G}^T_\alpha\Bigr) - 2\Bigr]^2 .
\label{eq:LossNormalization}    
\eeq
\item {\em Closure of the algebra}. This tests whether the set of candidate generators $\{\mathbb G\}$ forms a closed algebra 
$
\bigl[ \mathbb{G}_\alpha, \mathbb{G}_\beta\bigr] 
= \sum_{\gamma=1}^{N_g} a_{\alpha\beta}^{\gamma} \mathbb{G}_\gamma 
$
with some structure constants $a_{\alpha\beta}^{\gamma}$:
\beq
L_\text{closure} \bigl(\{\mathbb G\}, a_{\alpha\beta}^{\gamma} \bigr) = 
\sum_{ \alpha=1}^{N_g-1} \sum_{\substack{\beta=\alpha+1 }}^{N_g} 
\Tr \left(\mathbb{C}_{\alpha\beta}  \cdot\mathbb{C}_{\alpha\beta}^T\right),
\label{eq:LossClosure}
\eeq
where the closure mismatch is defined by
\beq 
\mathbb{C}_{\alpha\beta} \bigl(\{\mathbb G\},a_{\alpha\beta}^{\gamma}\bigr) \equiv
\bigl[ \mathbb{G}_\alpha, \mathbb{G}_\beta\bigr] 
- \sum_{\gamma=1}^{N_g} a_{\alpha\beta}^{\gamma} \mathbb{G}_\gamma .
\label{eq:closuremismatch}
\eeq
\end{enumerate}
The loss function in \cite{Forestano:2023fpj} was the sum of those four terms with suitable weight factors (hyperparameters) $h_\text{ortho}$, $h_\text{norm}$ and $h_\text{closure}$:
\beq
L_4 = 
 L_\text{inv} 
+h_\text{ortho}\,  L_\text{ortho}
+h_\text{norm}\, L_\text{norm}
+h_\text{closure}\, L_\text{closure}.
\label{eq:LossOldPaper}
\eeq

The analysis in \cite{Forestano:2023fpj} demonstrated that this method is capable of learning the symmetry groups (and their subgroups) generated by the oracles (\ref{oracle:O}) and (\ref{oracle:L}) without any prior assumptions, in a fully agnostic fashion. Since the method is completely general, however, the representations learned in \cite{Forestano:2023fpj} were not conveniently aligned with the axes in the feature space, thus differing from the nice conventional results found in the textbooks. To rectify this problem, here we introduce an additional loss term designed to encourage sparsity:
\beq
L_\text{sparsity} \bigl(\{\mathbb G\} \bigr) =
\sum_{ \alpha=1}^{N_g} 
\sum_{j=1}^n\sum_{k=1}^n
\sum_{\substack{j'=1}}^n\sum_{\substack{k'=1}}^n
\left|{\mathbb G}_\alpha^{(jk)}{\mathbb G}_\alpha^{(j'k')}\right| \Bigl(1-\delta_{jj'}\delta_{kk'}\Bigr),
\label{eq:LossSparsity}
\eeq
where ${\mathbb G}_\alpha^{(jk)}$ denotes the $jk$-component of $\mathbb G_\alpha$.
The full loss function is therefore
\beq
L\bigl(\{\mathbb G\}, a_{\alpha\beta}^{\gamma} \bigr)  = L_4\bigl(\{\mathbb G\}, a_{\alpha\beta}^{\gamma} \bigr) + h_\text{sparsity}\, L_\text{sparsity}\bigl(\{\mathbb G\} \bigr) 
\label{eq:LossThisPaper}
\eeq
and the minimization is performed over all generator components ${\mathbb G}_\alpha^{(jk)}$ and structure constants $a_{\alpha\beta}^{\gamma}$, using the {\sc Adam} optimizer with a learning rate between $0.0001$ and $0.01$. Unless otherwise specified, the hyperparameters $h$ were held fixed to 1. 

\section{The Lorentz group \texorpdfstring{$SO(1,3)$}{SO(1,3)}}
\label{sec:lorentz_group}

\begin{figure}[t]
    \centering
    \includegraphics[width=0.45\textwidth]{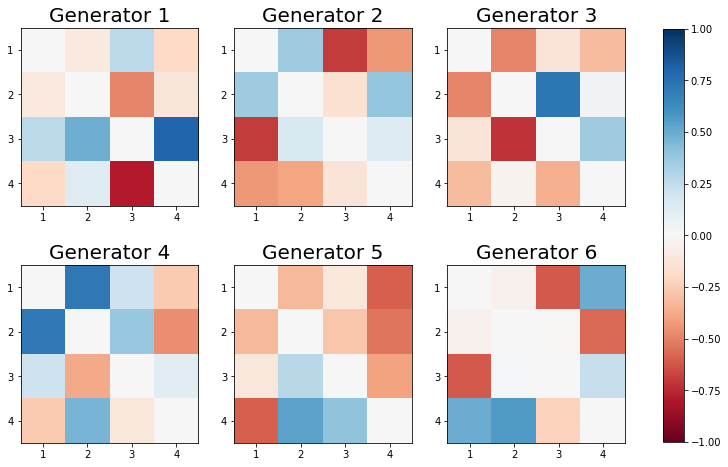}
    \includegraphics[width=0.45\textwidth]{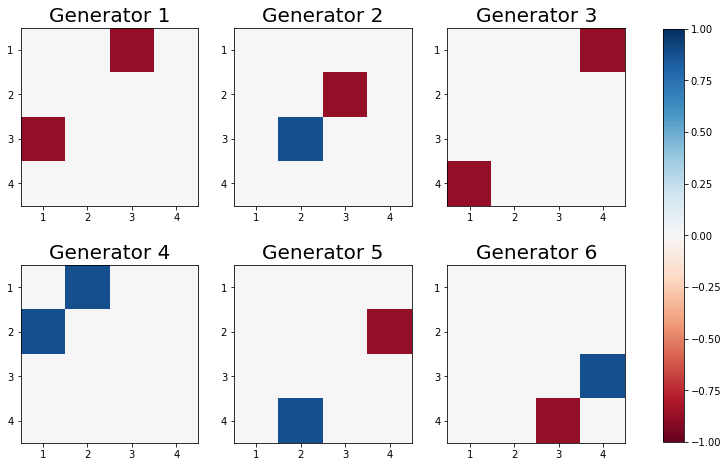}\\
    \includegraphics[width=0.3\textwidth]{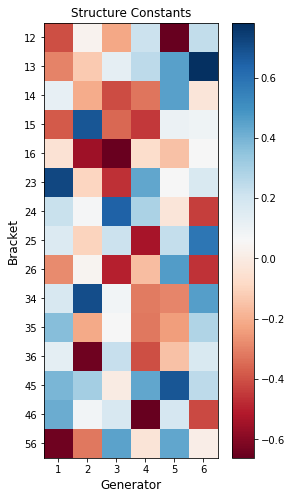}~~~~~~~~~~~~~~~~~~~~
    \includegraphics[width=0.3\textwidth]{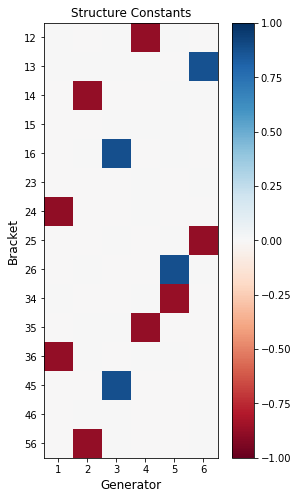}
    \caption{Left: the Lorentz group generators found in Ref.~\cite{Forestano:2023fpj} by using the loss function (\ref{eq:LossOldPaper}) and the oracle (\ref{oracle:L}) (top panels) and the corresponding structure constants (bottom panel). Right: the corresponding results after inclusion of the sparsity term (\ref{eq:LossSparsity}) with $h_\text{sparsity}=1$.  } \label{fig:Lor_6generators_not_sparse}
\end{figure}

As a reference benchmark, the left column of Figure~\ref{fig:Lor_6generators_not_sparse} shows the result from \cite{Forestano:2023fpj} for the $N_g=6$ generators of the Lorentz group, using the loss function (\ref{eq:LossOldPaper}) and the oracle (\ref{oracle:L}). In this and all subsequent such figures, each $n\times n$ panel represents a learned generator ${\mathbb J}_\alpha$ in matrix form. The values of the individual matrix elements are indicated by the color bar. In the bottom panels of Figure~\ref{fig:Lor_6generators_not_sparse}, each row (labeled $\alpha\beta=12,13,\ldots,56$) represents one of the 15 possible unique commutators $[\mathbb J_\alpha,\mathbb J_\beta]$, whereas the columns (labeled $\gamma=1,2,\ldots, 6$) represent the found generators $\mathbb J_\gamma$. Each cell then represents a structure constant $a_{\alpha\beta}^{\gamma}$, with its value given by the color bar.

While the six generators found in \cite{Forestano:2023fpj} do satisfy the four criteria 1-4 discussed in the previous section, they are not easily interpretable, since the generators end up being generic linear combinations of the familiar generators of boosts ${\mathbb K}_k$ and rotations ${\mathbb L}_k$, $k=1,2,3$. In other words, the representations in the top left panels in Figure~\ref{fig:Lor_6generators_not_sparse} are not sparse. After including the sparsity term (\ref{eq:LossSparsity}) in the loss function (\ref{eq:LossThisPaper}), we obtain the corresponding result shown in the right panels of Figure~\ref{fig:Lor_6generators_not_sparse}. We easily recognize $\mathbb J_1=-\mathbb K_2$, $\mathbb J_3=-\mathbb K_3$ and $\mathbb J_4=\mathbb K_1$ as the canonical boost generators and $\mathbb J_2=\mathbb L_3$, $\mathbb J_5=-\mathbb L_2$ and $\mathbb J_6=-\mathbb L_1$ as the canonical generators of rotations.

\begin{figure}[tbp]
    \centering
    \includegraphics[width=0.65\textwidth]{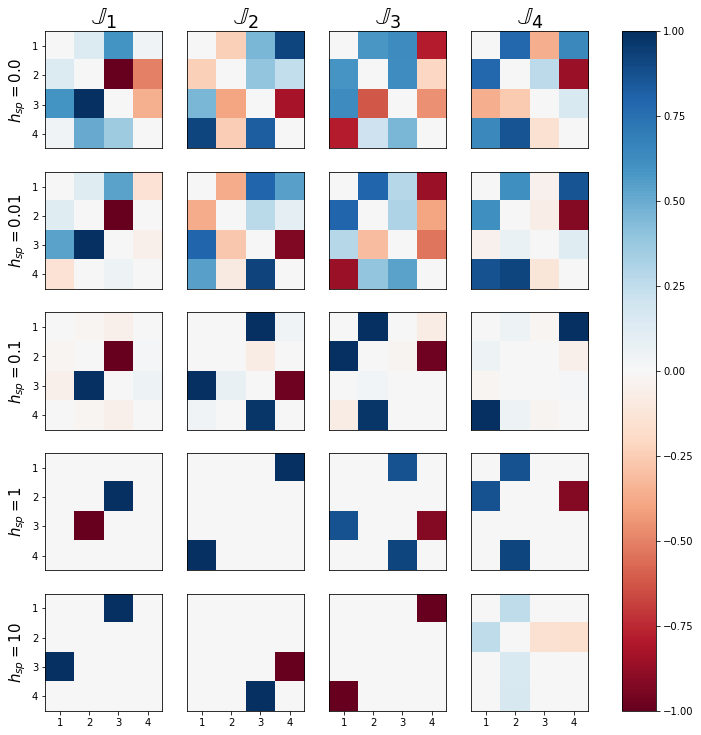}
    \caption{The learned generators of the $N_g=4$ subalgebra of the Lorentz group for different values of the $h_\text{sparsity}$ hyperparameter: $h_\text{sparsity}=0, 0.01, 0.1, 1, 10$ (from top to bottom). Each training session ran for 3,000 epochs. } \label{fig:Lor_4generators}
\end{figure}

Like any ML approach, our method relies on the minimization of a loss function, which involves some hyperparameter tuning. For optimal results, we do need to successfully balance the relative weights of the different terms in the loss function (\ref{eq:LossThisPaper}). Figure~\ref{fig:Lor_4generators} illustrates the dependence of the learned generators for the $N_g=4$ subalgebra of the Lorentz group on the sparsity hyperparameter $h_\text{sparsity}$, which is scanned from 0 to 10. As the sparsity term is gradually turned on, the learned representations become sparser, and for a certain range of values of $h_\text{sparsity}$, we recover a canonical set $\mathbb J_1=\mathbb L_3$, $\mathbb J_2=\mathbb K_2 + \mathbb L_1$, $\mathbb J_3=\mathbb K_1 - \mathbb L_2$, and $\mathbb J_4=\mathbb K_3$. However, increasing $h_\text{sparsity}$ further leads to solutions which are very sparse, but only at the cost of violating one of the other conditions, in this case normalization --- note that $\mathbb J_4$ in the last row of Figure~\ref{fig:Lor_4generators} is essentially trivial.

\section{The \texorpdfstring{$U(n)$}{U(n)} and \texorpdfstring{$SU(n)$}{SU(n)} family}
\label{sec:SUN}

We now turn our attention to the $U(n)$ and $SU(n)$ families which were not considered in \cite{Forestano:2023fpj} or elsewhere in the current context. The two families are related as $U(n)=SU(n)\times U(1)$, where the $U(1)$ factor accounts for the $\det U=1$ condition. Since the invariant polynomial (\ref{oracle:U}) is defined over the field of complex numbers, we need to make a few adjustments to the procedure in Section~\ref{sec:loss}: i) create a dataset $\{\mathbf x\}$ which now consists of $m$ $n$-dimensional complex vectors; ii) use the complex form of the functions defined in the loss function by replacing any products of $\mathbb G \mathbb G$ with $\mathbb G \mathbb G^\ast$; and iii) insert the conventional imaginary unit factor in front of $\varepsilon$ in (\ref{eq:deltaf}) as $\delta{\f} = \mathbb{I} + i \varepsilon \, {\mathbb G}$. Upon training with $n^2$ generators, we will find the algebra of $U(n)$. In order to find the algebra of $SU(n)$, we take advantage of the ability of our method to find sets of generators forming closed algebras. Therefore, all we need to do is reduce the number of generators to $n^2-1$ and proceed as before.

\subsection{The case of \texorpdfstring{$U(2)$}{U(2)} and \texorpdfstring{$SU(2)$}{SU(2)}}

\begin{figure}[t]
    \centering
    \includegraphics[width=0.48\textwidth]{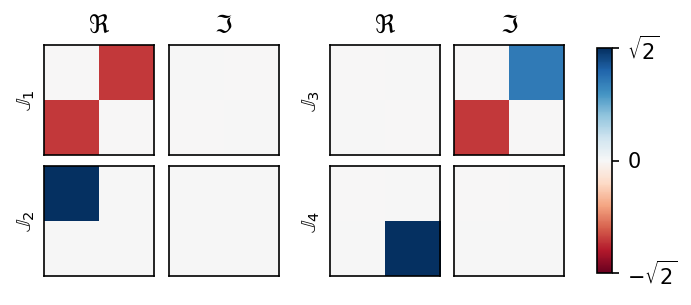}
    \includegraphics[width=0.48\textwidth]{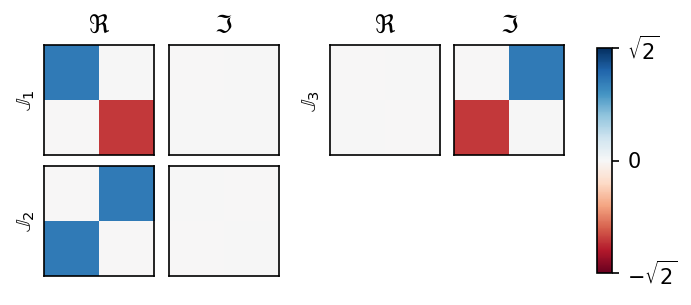}
    \caption{The learned sparse generators for the case of $U(2)$ (left) and $SU(2)$ (right). } \label{fig:SU2}
\end{figure}

Our results for the case of $n=2$ are shown in Figure~\ref{fig:SU2}. Since the components of the generator matrices are now complex, we show separate panels for their real ($\mathfrak R$) and imaginary ($\mathfrak I$) parts. The generators of $U(2)$ are comprised of the generators of $SU(2)$, usually taken as the Pauli matrices 
\beq
\sigma_1=
\begin{pmatrix}
0 & 1 \\ 1 & 0
\end{pmatrix}, \qquad
\sigma_2=
\begin{pmatrix}
0 & -i \\ i & 0
\end{pmatrix}, \qquad
\sigma_3=
\begin{pmatrix}
1 & 0 \\ 0 & -1
\end{pmatrix},
\label{Pauli}
\eeq
supplemented with the $2\times 2$ identity matrix $\sigma_0={\mathbb I}_2$. We see that for the $N_g=4$ case of $U(2)$ displayed on the left side of Figure~\ref{fig:SU2}, the method correctly finds $\mathbb J_1=-\sigma_1$, $\mathbb J_3=-\sigma_2$, $\mathbb J_2=(\sigma_0+\sigma_3)/\sqrt{2}$, $\mathbb J_4=(\sigma_0-\sigma_3)/\sqrt{2}$. The last two linear combinations are chosen by the algorithm because they are sparser than the canonical set $\sigma_0$ and $\sigma_3$. Note that when a generator matrix has a single nonvanishing element, its value is fixed to $\pm\sqrt{2}$ by the normalization condition (\ref{eq:LossNormalization}). When repeating the exercise with $N_g=3$, as is the case of $SU(2)$, we find the result on the right in Figure~\ref{fig:SU2}. We recognize $\mathbb J_1=\sigma_3$, $\mathbb J_2=\sigma_1$, $\mathbb J_3=-\sigma_2$, which is precisely the canonical $SU(2)$ algebra up to signs and relabelling.

\subsection{The case of \texorpdfstring{$U(3)$}{U(3)} and \texorpdfstring{$SU(3)$}{SU(3)}}

\begin{figure}[t]
    \centering
    \includegraphics[width=0.75\textwidth]{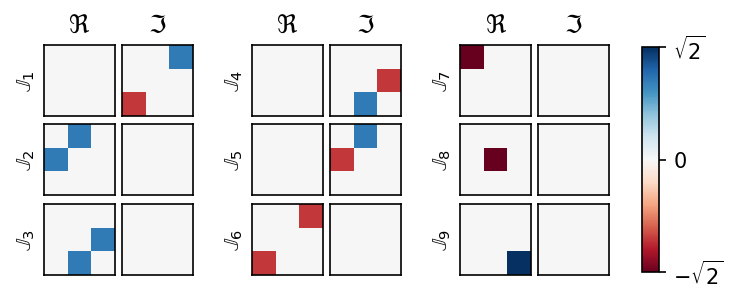}\\
    \includegraphics[width=0.75\textwidth]{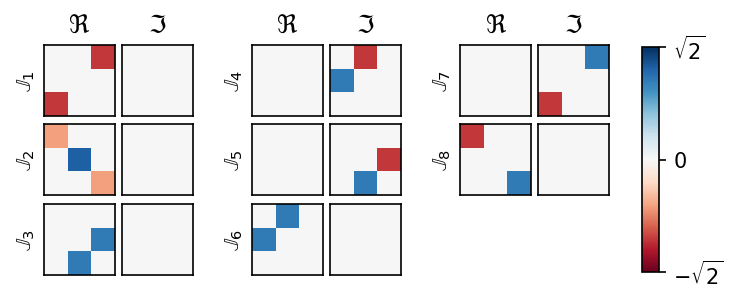}
    \caption{The same as Figure~\ref{fig:SU2}, but for the case of $U(3)$ (top panels) and $SU(3)$ (bottom panels). } \label{fig:SU3}
\end{figure}

We now move on to the case of $n=3$. The generators of $SU(3)$ are the Gell-Mann matrices which are the $SU(3)$ analogues of the Pauli matrices (\ref{Pauli})
\begin{align}
\lambda_1 &=
\begin{pmatrix}
0 & 1 & 0\\ 1 & 0 & 0 \\ 0 & 0 & 0
\end{pmatrix}, \qquad
\lambda_2 =
\begin{pmatrix}
0 & -i & 0\\ i & 0 & 0 \\ 0 & 0 & 0
\end{pmatrix}, \qquad
\lambda_3 =
\begin{pmatrix}
1 & 0 & 0\\ 0 & -1 & 0 \\ 0 & 0 & 0
\end{pmatrix}, \notag\\
\lambda_4 &=
\begin{pmatrix}
0 & 0 & 1\\ 0 & 0 & 0 \\ 1 & 0 & 0
\end{pmatrix}, \qquad
\lambda_5 =
\begin{pmatrix}
0 & 0 & -i\\ 0 & 0 & 0 \\ i & 0 & 0
\end{pmatrix}, \label{Gell-Mann}\\
\lambda_6 &=
\begin{pmatrix}
0 & 0 & 0\\ 0 & 0 & 1 \\ 0 & 1 & 0
\end{pmatrix}, \qquad
\lambda_7 =
\begin{pmatrix}
0 & 0 & 0\\ 0 & 0 & -i \\ 0 & i & 0
\end{pmatrix}, \qquad
\lambda_8 = \frac{1}{\sqrt{3}}
\begin{pmatrix}
1 & 0 & 0\\ 0 & 1 & 0 \\ 0 & 0 & -2
\end{pmatrix}.\notag
\end{align}
Our results are shown in Figure~\ref{fig:SU3}. Upon inspection, we see that the generators are successfully reproduced. In particular, for $SU(3)$, the algorithm found
$\mathbb J_1=-\lambda_4$,
$\mathbb J_2=(\lambda_8 - \sqrt{3}\lambda_3)/2$,
$\mathbb J_3=\lambda_6$,
$\mathbb J_4=\lambda_2$,
$\mathbb J_5=\lambda_7$,
$\mathbb J_6=\lambda_1$,
$\mathbb J_7=-\lambda_5$,
$\mathbb J_8=-(\lambda_3+\sqrt{3}\lambda_8)/2$.

\subsection{The case of \texorpdfstring{$U(4)$}{U(4)} and \texorpdfstring{$SU(4)$}{SU(4)}}

\begin{figure}[t]
    \centering
    \includegraphics[width=0.85\textwidth]{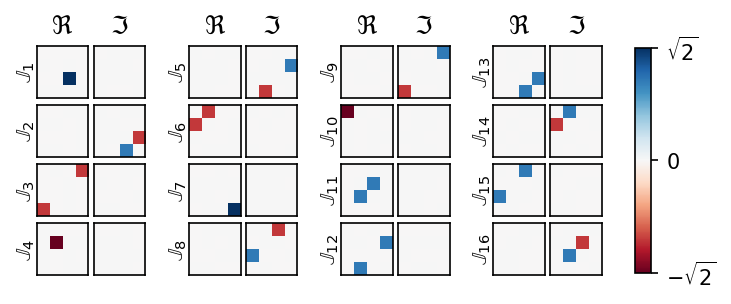}\\
    \includegraphics[width=0.85\textwidth]{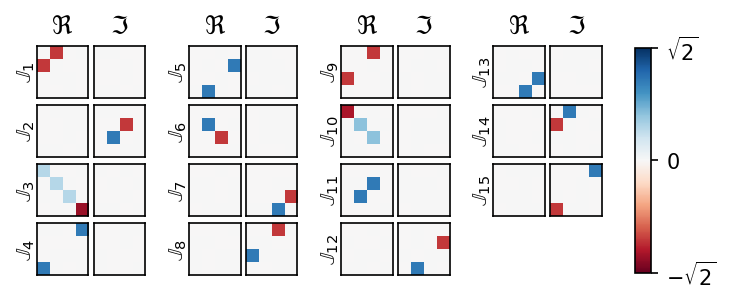}
    \caption{The same as Figure~\ref{fig:SU3}, but for the case of $U(4)$ (top panels) and $SU(4)$ (bottom panels). } \label{fig:SU4}
\end{figure}

Figure \ref{fig:SU4} shows the results for the next most complicated case, $n=4$. The generators of $SU(4)$ are the corresponding generalizations of the Gell-Mann matrices (\ref{Gell-Mann}) and consist of 6 symmetric matrices with real off-diagonal components, 6 anti-symmetric matrices with purely imaginary off-diagonal components, and 3 matrices with real diagonal components. This is precisely what we observe in the lower panels of Figure \ref{fig:SU4}. The $U(4)$ algebra has one additional diagonal generator, which can be taken as the identity matrix $\mathbb I_4$. Once again, the sparsity term in the loss function forces the $U(4)$ generators with real diagonal components to have a single non-zero entry, see $\mathbb J_1$, $\mathbb J_4$, $\mathbb J_7$, and $\mathbb J_{10}$.

\subsection{The case of \texorpdfstring{$SU(5)$}{SU(5)} and beyond}

\begin{figure}[tbp]
    \centering
    \includegraphics[width=0.85\textwidth]{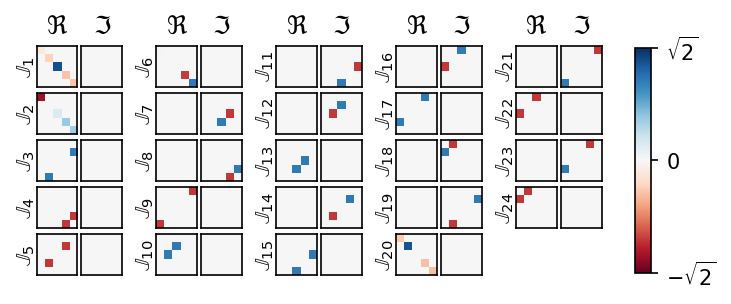}\\
    \includegraphics[width=0.85\textwidth]{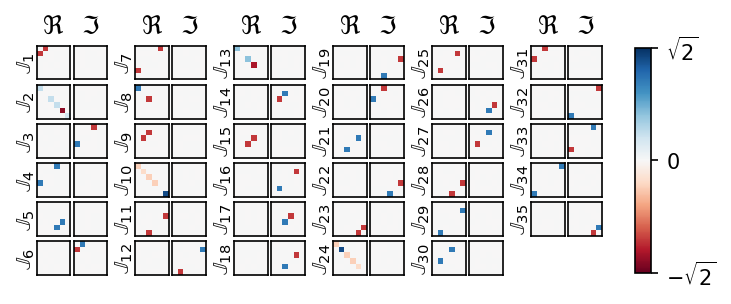}\\
    \caption{The learned sparse generators for the case of $SU(5)$ (top panels) and $SU(6)$ (bottom panels). } \label{fig:SU56}
\end{figure}

We now turn to the cases of $n=5$ and $n=6$. The learned generators for $SU(5)$ and $SU(6)$ are shown in Figure~\ref{fig:SU56}. In the case of $SU(5)$, we have 4 generators with real diagonal entries ($\mathbb J_1$, $\mathbb J_2$, $\mathbb J_6$ and $\mathbb J_{20}$), 10 symmetric matrices with real off-diagonal components
($\mathbb J_\alpha$ with $\alpha\in\{3,4,5,9,10,13,15,17,22,24\}$),
and 10 anti-symmetric matrices with imaginary off-diagonal components
($\mathbb J_\alpha$ with $\alpha\in\{7,8,11,12,14,16,18,19,21,23\}$).
In the case of $SU(6)$, out of the 35 total generators, we have 5 generators with real diagonal entries 
($\mathbb J_\alpha$ with $\alpha\in\{2,8,10,13,24\}$),
15 symmetric matrices with real off-diagonal components
($\mathbb J_\alpha$ with $\alpha\in\{1,4,5,7,9,11,15,21,23,25,28,29,30,31,34\}$),
and 15 anti-symmetric matrices with imaginary off-diagonal components ($\mathbb J_\alpha$ with $\alpha\in\{3,6,12,14,16,17,18,19,20,22,26,27,32,33,35\}$).
Having demonstrated the successful discovery of the $SU(5)$ and $SU(6)$ generators in Figure~\ref{fig:SU56}, we believe that our procedure has proven to be fully generalizable for any higher dimensional groups $SU(n)$ with $n>6$.

\section{Summary and Outlook}
\label{sec:summary}

In this letter we demonstrated how to use machine learning to derive sparse representations of the generators of an arbitrary Lie algebra. The key observation is that a Lie group preserves a certain invariant polynomial under the group action on the vector space (for the examples considered here, those were the oracle functions (\ref{oracle:O}-\ref{oracle:U})). As shown in \cite{Forestano:2023fpj}, one can design suitable loss functions for the generator components, so that minimization of the loss function leads to a set of valid generators. In this letter we extended the approach of \cite{Forestano:2023fpj} by adding terms to the loss function that encourage finding sparse representations. We validated our method for the Lorentz group (in Section~\ref{sec:lorentz_group}) and for the $U(n)$ and $SU(n)$ family of Lie groups (in Section~\ref{sec:SUN}). Although not shown here, we have also verified the procedure on all of the $SO(n)$ examples considered in \cite{Forestano:2023fpj}.

We should note that all of our numerical work was done on a personal laptop, and did not use any high performance facilities. All models were trained for 5,000 epochs and the typical training times ranged from a few minutes to a couple of hours for the case of $SU(6)$. While the method is applicable to arbitrarily large groups, in practice the training in such cases may become relatively slow and could benefit from parallelization, further code optimization, further hyperparameter tuning, and better understanding of the loss landscape. Examples of much larger groups, like the exceptional Lie groups \cite{Ramond:1976aw}, will be explored in future work.

\acknowledgements
We thank A.~Davis, S.~Gleyzer, R.~Houtz, K.~Kong, S.~Mrenna, H.~Prosper and P.~Shyamsundar for useful discussions. We thank P.~Ramond for group theory insights and inspiration.  This work is supported in part by the U.S.~Department of Energy award number DE-SC0022148.

\bibliography{references}

\end{document}